\documentclass[%
reprint,
amsmath,amssymb,
aps,
prb,
superscriptaddress
]{revtex4-1}

\usepackage[dvipsnames]{xcolor}
\usepackage{graphicx}
\usepackage{dcolumn}
\usepackage{bm}

\newcommand{\vect}[1]{\boldsymbol{#1}}

\begin{document}

\preprint{APS/123-QED}

\title{Third-order intrinsic anomalous Hall effect with generalized semiclassical theory}
\author{Longjun Xiang}
\affiliation{College of Physics and Optoelectronic Engineering, Shenzhen University,
Shenzhen 518060, China}
\author{Chao Zhang}
\affiliation{Department of Physics,
The Hong Kong University of Science and Technology, Clear Water Bay, Kowloon, Hong Kong, China}
\author{Luyang Wang}
\affiliation{College of Physics and Optoelectronic Engineering, Shenzhen University,
Shenzhen 518060, China}
\author{Jian Wang}
\email{jianwang@hku.hk}
\affiliation{College of Physics and Optoelectronic Engineering, Shenzhen University,
Shenzhen 518060, China}
\affiliation{Department of Physics, The University of Hong Kong, Pokfulam Road, Hong Kong, China}
\date{\today}

\begin{abstract}
The linear intrinsic anomalous Hall effect (IAHE) and second-order IAHE have been intensively investigated
in time-reversal broken systems. However, as one of the important members of the nonlinear Hall family,
the investigation of third-order IAHE remains absent due to the lack of an appropriate theoretical approach,
although the third-order extrinsic AHE has been studied within the framework of first- and
second-order semiclassical theory. Herein, we generalize the semiclassical theory
for Bloch electrons under the uniform electric field up to the third order using the wavepacket method
and based on which we predict that the third-order IAHE can also occur in time-reversal broken systems.
Same as the second-order IAHE, we find the band geometric quantity, the second-order field-dependent
Berry curvature arising from the second-order field-induced \textit{positional shift},
plays a pivotal role to observe this effect.
Moreover, with symmetry analysis, we find that the third-order IAHE, as the leading contribution,
is supported by $15$ time-reversal broken 3D magnetic point groups (MPGs), corresponding to a wide class
of antiferromagnetic (AFM) materials. 
Guided by the symmetry arguments, a two-band model is chosen to demonstrate the generalized theory.
Furthermore, the generalized third-order semiclassical theory
depends only on the properties of Bloch bands,
implying that it can also be employed to explore the IAHE in realistic AFM materials,
by combining with first-principles calculations.
\end{abstract}

\maketitle

\textit{Introduction}.--- As one of the important phenomena in condensed matter physics,
the charge Hall effect and its variants,
such as the anomalous Hall effect \cite{Karplus, RevModPhys.82.1539}
and spin Hall effect \cite{SCZhang, Sinova2004},
have been studied continuously since its discovery in 1879 \cite{Hall}.
For a long time, it has been believed that the observation
of charge Hall current requires broken time-reversal ($\mathcal{T}$) symmetry
\cite{landau1999statistical}.
However, within the framework of the first-order semiclassical theory \cite{RevModPhys.82.1959},
Sodemann and Fu predicted \cite{PhysRevLett.115.216806}
that the second-order nonlinear Hall effect (NLHE) can exist even in $\mathcal{T}$-invariant materials,
driven by the band geometric quantity Berry curvature dipole (BCD).
Recently, this extrinsic (proportional to the relaxation time $\tau$)
NLHE has been observed experimentally\cite{Nature2019, Raffaele} and
investigated theoretically\cite{Kamp,NRP2021}.

In fact, slightly before investigating NLHEs within $\mathcal{T}$-invariant materials,
the second-order nonlinear intrinsic (free of $\tau$ and depends only on the band topology)
anomalous Hall effect (IAHE) in magnetic systems
has been predicted by Y. Gao \textit{et al.}\cite{PhysRevLett.112.166601},
by extending the semiclassial theory up to the second order.
Very recently, with this second-order semiclassical theory,
two research groups further predicted that the second-order nonlinear IAHE
can be observed in $\mathcal{PT}$-symmetric compensated antiferromagnetic (AFM) materials CuMnAs \cite{wang2021intrinsic}
and Mn$_2$Au \cite{liu2021intrinsic}.
Additionally, the intrinsic nonlinear planar Hall effects
also have been proposed in magnetic Weyl semimetals \cite{LPHE} and
in nonmagnetic polar and chiral crystals\cite{NPHE}, respectively.
Furthermore, using the second-order semiclassical theory,
the extrinsic third-order NLHE has been investigated in materials
\cite{lai2021third, PhysRevB.105.045118, TaIrTe4, Nandy, ZhiMinLiao} with $\mathcal{T}$ and $\mathcal{P}$ symmetries,
suggesting that the third-order Hall signal can be dominant in some materials.
Besides, the extrinsic third-order nonlinear AHE, induced by Berry curvature quadrupole, 
can also be the leading contribution in some $\mathcal{T}$-broken materials,
predicted by Zhang \textit{et al.} \cite{zhang2020higherorder} with the first-order semiclassical theory.
However, as one of the important members of the nonlinear intrinsic anomalous Hall family,
the third-order IAHE remains unexplored so far.

In this work, we generalize the semiclassical theory for Bloch electrons
under the uniform electric field to third order using the wavepacket method
and based on this we predict the existence of third-order IAHE in $\mathcal{T}$-broken systems.
Like the second-order semiclassical theory \cite{PhysRevLett.112.166601},
we find the band geometric quantity, the second-order field dependent Berry curvature (BC) arising from
the second-order field-induced \textit{positional shift} \cite{PhysRevLett.112.166601},
plays a key role in observing this effect.
Moreover, from symmetry analysis, we find that the third-order IAHE, as the leading contribution, 
can be hosted by $15$ $\mathcal{T}$-broken 3D magnetic point groups (MPGs),
corresponding to a wide class of AFM materials.
Following the symmetry arguments,
a two-band model is chosen to illustrate our generalized theory.

\bigskip
\textit{The third-order semiclassical theory}.---
Following the spirit of the semiclassical wavepacket approach \cite{RevModPhys.82.1959, PhysRevLett.112.166601},
we focus on the $0$th band (Throughout this work, we will only consider the Abelian case)
and first construct the wavepacket \cite{PhysRevLett.112.166601} as follows:
\begin{align}
|\Psi \rangle 
&= 
\int_{\vect{p}}  e^{i \vect{p} \cdot \vect{r}}
\left( C_0(\vect{p}) |u_0(\vect{p} )\rangle 
+
\sum_{n \ne 0} C_n(\vect{p})|u_n(\vect{p} )\rangle
\right)
\label{wavepacket}
\end{align}
where $|u_n(\vect{p})\rangle$ is the Bloch state,
$C_0$ is the zeroth-order amplitude with $|C_0|^2=\delta(\vect{p}-\vect{p}_c)$
to normalize the wavepacket up to the first order\cite{RevModPhys.82.1959},
$C_n=C_n^{(1)}+C_n^{(2)}$ with $C_n^{(1)}$/$C^{(2)}_n$ the first-order/second-order amplitude
in terms of the external electric field $\vect{E}$. Note that
both $C_n^{(1)}$ and $C^{(2)}_n$ are related to $C_0$, for example \cite{PhysRevLett.112.166601},
\begin{align}
C_n^{(1)} & = \mathcal{M}_{n0} C_0 
\label{C1C0}
\end{align}
where $\mathcal{M}_{n0}=\vect{E}\cdot\vect{\mathcal{A}}_{n0}/(\epsilon_0-\epsilon_n)$
with $\vect{\mathcal{A}}_{n0}=\langle u_n|i\partial_{\vect{k}_c}|u_0\rangle$ the 
interband Berry connection and $\epsilon_{0/n}$ the unperturbed band energy.
Note that Eq.(\ref{C1C0}) is obtained by solving the time-dependent Schr\"{o}dinger equation,
therefore the relation between $C_n^{(2)}$ and $C_0$ can be identified similarly \cite{supplemental}:
\begin{align}
C^{(2)}_n
&=
\left(
\sum_{m \ne 0}
\frac{\vect{E}\cdot\vect{\mathcal{A}}_{nm}\mathcal{M}_{m0}}{\epsilon_0-\epsilon_n}
+
\frac{\vect{E} \cdot i\partial_{\vect{p}} \mathcal{M}_{n0}}{\epsilon_0-\epsilon_n}
\right)C_0
\nonumber \\
&+
\dfrac{\vect{E} \cdot [(i\partial_{\vect{p}} - \vect{r}_c) C_0]\mathcal{M}_{n0}}{\epsilon_0-\epsilon_n} 
\label{C2}
\end{align}
For the second order, we also must include a second-order correction
$\dfrac{1}{2}\sum_{m \neq 0} \mathcal{M}_{0m} \mathcal{M}_{m0} C_0$
for $C_0$ in the first term of Eq.(\ref{wavepacket}) to normalize
the wavepacket up to the second order\cite{energy}, which, together with Eq. (\ref{C2}),
complete the construction of the second-order wavepacket.

Once the wavepacket is constructed,
one can immediately calculate the wavepacket Lagrangian \cite{PhysRevLett.112.166601}
and then derive the equation of motion (EOM) describing the dynamics of the momentum ($\vect{k}_c$)
and position ($\vect{r}_c$) centers of wavepacket under the uniform electric field
($e=\hbar=1$):
\begin{align}
\dot{\vect{r}}_c &=\partial_{\vect{k}_c} \bar{\epsilon} -\dot{\vect{k}}_c \times \bar{\vect{\Omega}}, \label{EOM1} \\
\dot{\vect{k}}_c &=-\vect{E},
\label{EOM2}
\end{align}
where $\partial_{\vect{k}_c}\equiv \partial/\partial \vect{k}_c$.
Interestingly, we find that both the velocity and force equations keep the same form
as the first- and second-order semiclassical EOM, but the band energy \cite{energy, Xiaocong1, Xiaocong2}
and Berry curvature\cite{PhysRevLett.112.166601} in Eq.(\ref{EOM1}) should include
the correction from the external electric field.
Particularly, in Eq.(\ref{EOM1}), $\bar{\vect{\Omega}} \equiv \vect{\Omega}+\vect{\Omega}^{(1)}+\vect{\Omega}^{(2)}$
is the BC accurate to second order in terms of $\vect{E}$,
where $\vect{\Omega}$ is the conventional BC
and $\vect{\Omega}^{(1/2)}$ are the first- and second-order field-dependent BC, respectively.
By calculating the first-order field-induced \textit{positional shift}
for the position center $\vect{r}_c$ with the constructed wavepacket,
Y. Gao \textit{et al.} \cite{PhysRevLett.112.166601}
have successfully developed the second-order semiclassical theory.
Herein, we further derive the second-order
\textit{positional shift} to generalize the semiclassical theory up to the third order.

Within the constructed wavepacket, the position center $\vect{r}_c$ can be expressed as \cite{GaoYangLT}:
\begin{align}
\vect{r}_c 
\equiv
\langle \Psi | \vect{r} | \Psi\rangle
=
\partial_{\vect{k}_c} \gamma
+
\vect{\mathcal{A}}_{0}
+
\vect{\mathcal{A}}_{0}^{(1)}
+
\vect{\mathcal{A}}_{0}^{(2)}
\end{align}
where $\gamma = -\text{arg}(C_0)$, $\vect{\mathcal{A}}_0 = \langle u_0|i\partial_{\vect{k}_c}|u_0\rangle$
is the intraband Berry connection, 
$\vect{\mathcal{A}}^{(1)}_0=2\text{Re} \sum_{n \neq 0}\vect{\mathcal{A}}_{0n}\mathcal{M}_{n0}$
is the first-order \textit{positional shift}, firstly derived by Y. Gao \textit{et al.}\cite{PhysRevLett.112.166601}.
At this stage, by taking the curl of $\vect{\mathcal{A}}_{0}$ and $\vect{\mathcal{A}}_{0}^{(1)}$,
the conventional BC and the first-order field-dependent BC $\vect{\Omega}^{(1)}$
can be obtained and hence the first- and second-order semiclassical theories are formulated.
Furthermore, with the second-order wavepacket constructed in Eq.(\ref{wavepacket}), we obtain
the second-order \textit{positional shift} \cite{supplemental}:
\begin{align}
\vect{\mathcal{A}}_0^{(2)}
&=
\text{Re}
\sum_{n \neq 0}^{m \neq 0}
\left[
\dfrac{2\vect{\mathcal{A}}_{0n}\vect{E}\cdot\vect{\mathcal{A}}_{nm} \mathcal{M}_{m0}}{\epsilon_{0}-\epsilon_{n}}
-
\mathcal{M}_{0n}  \vect{\mathcal{A}}_{nm} \mathcal{M}_{m0}
\right]
\nonumber \\
&-
\text{Re} 
\sum_{n \neq 0}
\left[
\dfrac{2\vect{E}\cdot\vect{\mathcal{A}}_{0} \mathcal{M}_{n0} \vect{\mathcal{A}}_{0n} }
{\epsilon_{0}-\epsilon_{n}}
-
\vect{\mathcal{A}}_{0} \mathcal{M}_{0n} \mathcal{M}_{n0}
\right]
\nonumber \\
&-
\text{Re}
\sum_{n \neq 0}
\left[
\mathcal{M}_{0n}i\partial_{\vect{k}_c} \mathcal{M}_{n0}
-
\dfrac{\vect{\mathcal{A}}_{0n} \vect{E} \cdot i\partial_{\vect{k}_c} \mathcal{M}_{n0}}
{\epsilon_{0}-\epsilon_{n}}
\right]
\nonumber \\
&-
\text{Re}
\sum_{n \neq 0}
\mathcal{M}_{n0}
\vect{E} \cdot i\partial_{\vect{k}_c} 
\left( 
\dfrac{
\vect{\mathcal{A}}_{0n}}
{\epsilon_{0}-\epsilon_{n}}
\right)
\end{align}
where we have replaced the momentum center $\vect{p}_c$ with $\vect{k}_c$ at the final results.
Importantly, by defining
$\mathcal{A}^{\alpha,(2)}_{0} \equiv T^0_{\alpha\beta\gamma} E_\beta E_\gamma$
with $T^0_{\alpha\beta\gamma}$ the second-order
\textit{Berry-connection polarizability tensor} (BPT) for band $0$,
we find
\begin{align}
T^0_{\alpha\beta\gamma}
&=
\text{Re}
\sum_{n}
\left(
\mathcal{U}_{0n}^{\alpha\beta\gamma}
+
\mathcal{U}_{0n}^{\beta\alpha\gamma}
-
\mathcal{U}_{n0}^{\beta\gamma\alpha}
-
\sum_{m}
\mathcal{V}_{0nm}^{\alpha\beta\gamma}
\right)
\end{align}
with
\begin{align*}
\mathcal{U}_{0n}^{\alpha\beta\gamma}
&\equiv
M_{0n}^\beta 
\left(
\mathcal{A}_{0}^\alpha 
-
\mathcal{A}_{n}^\alpha 
-
i\partial_\alpha
\right)
M_{n0}^\gamma,
\\
\mathcal{V}_{0nm}^{\alpha\beta\gamma}
&\equiv
\left(
2M_{0n}^\alpha \mathcal{A}_{nm}^\beta 
+
M_{0n}^\beta \mathcal{A}_{nm}^\alpha 
\right)
M_{m0}^\gamma 
\bar{\delta}_{nm},
\end{align*}
where $\mathcal{A}^\alpha_n=\langle u_n|i\partial_\alpha|u_n\rangle$
with $\partial_\alpha \equiv \partial/\partial k_c^\alpha$
and $M_{mn}^\alpha=\mathcal{A}^\alpha_{mn}/(\epsilon_n-\epsilon_m)\bar{\delta}_{nm}$ with
$\bar{\delta}_{nm} \equiv 1-\delta_{nm}$.
Interestingly, under $U(1)$ gauge transformation $|u_n\rangle \rightarrow e^{i\phi_n}|u_n\rangle$, 
we find that $\mathcal{A}_n^\alpha \rightarrow \mathcal{A}_n^\alpha-\partial_\alpha \phi_n$,
$\mathcal{A}^\alpha_{mn} \rightarrow e^{i(\phi_n-\phi_m)} \mathcal{A}_{mn}^\alpha$,
and $M_{mn}^\alpha \rightarrow e^{i(\phi_n-\phi_m)} M_{mn}^\alpha$,
therefore, $\mathcal{U}, \mathcal{V}$, and $T$
are $U(1)$ gauge invariant.
Same as the physical implication of $\vect{\mathcal{A}}_{0}^{(1)}$\cite{PhysRevLett.112.166601},
$\vect{\mathcal{A}}_{0}^{(2)}$ stands for a second-order correction to
the conventional Berry connection $\vect{\mathcal{A}}_0$
of the unperturbed band,
which means that the wavepacket also acquires a shift
$\vect{\mathcal{A}}_{0}^{(2)}$ in its center $\vect{r}_c$ of mass position.
In addition, we note that $\vect{\mathcal{A}}^{(2)}_{0}$ also
respects the periodicity of the lattice due to the uniform external field,
hence it does not cause any macroscopic charge density gradient and
also will not affect the electron chemical potential profile \cite{PhysRevLett.112.166601, GaoYang2019}.
The second-order BPT is the central concept of our third-order semiclassical theory.

By taking the curl of the second-order field-dependent Berry connection,
we obtain the field-dependent BC $\vect{\Omega}^{(2)}$ at the same order.
Furthermore, substituting $\vect{\Omega}^{(2)}$ into Eq.(\ref{EOM1}), the third-order
semiclassical theory is established when the band energy $\bar{\epsilon}$ is corrected to third-order\cite{supplemental}.
Like the second-order semiclassical theory,
the third-order EOM and the second-order field-dependent BC,
plays an essential role in investigating the IAHE in AFM materials, as will be illustrated below,
especially when the linear and second-order IAHE signals are forbidden by symmetry.
The third-order semiclassical theory, with the second-order field-dependent BC originating from the
second-order field-induced \textit{positional shift} as the band geometric quantity,
is our first main result.

\bigskip
\textit{The third-order IAHE}. ---
Within the framework of third-order semiclassical transport theory,
if we ignore the scattering effects arising from impurities,
the third-order intrinsic Hall current density can be expressed as
{$J_{\alpha}=\chi^{\text{int}}_{\alpha\beta\gamma\eta}E_\beta E_\gamma E_\eta$},
where
\begin{align}
\chi_{\alpha\beta\gamma\eta}^{\text{int}} = \int_k
\Lambda_{\alpha\beta\gamma\eta} f_0
\label{currentdensity}
\end{align}
is the third-order intrinsic Hall conductivity, which is a rank-$4$ tensor.
In Eq.(\ref{currentdensity}), $f_0$ is the equilibrium Fermi distribution function, and
\begin{align}
\Lambda_{\alpha\beta\gamma\eta}
=
\sum_n
\left[
\partial_\beta T_{\alpha\gamma\eta}^n
-
\partial_\alpha T_{\beta\gamma\eta}^n
\right]
\end{align}
is the integrand for the third-order intrinsic Hall conductivity,
which is antisymmetric in its first two indices and symmetric in its last two
indices, namely $\chi^{\text{int}}_{\alpha\beta\gamma\eta}=-\chi^{\text{int}}_{\beta\alpha\gamma\eta}$
and $\chi^{\text{int}}_{\alpha\beta\gamma\eta}=\chi^{\text{int}}_{\beta\alpha\eta\gamma}$.
By performing
an integration by parts for Eq.(\ref{currentdensity}),
we find that the third-order IAHE is also a Fermi liquid property
\cite{PhysRevLett.115.216806, PhysRevLett.112.166601, wang2021intrinsic, liu2021intrinsic, Haldane},
as expected.

\begin{center}
\begin{table}
\caption{\label{tab1} {The classification for $90$ $\mathcal{T}$-broken MPGs in 3D with the third-,
and fourth-order IAHEs, as the leading contribution.}}
\begin{tabular}{ p{3.5cm} p{5.0cm} }
\hline
\hline
\\ [0.01em]
The order of IAHE & The MPGs for IAHE \\ [0.25cm]
\hline
\\ [0.01em]
Third-order &
$mmm$, $4'/m$, $4/mmm$, $4'/mm'm$, $\bar{3}m$, $6'$, $6'/m'$, $6'22'$, $6'mm'$, $\bar{6}m2$
$6/mmm$, $6'/m'mm'$, $m\bar{3}$, $4'32'$, $m\bar{3}m'$
\\ [0.20em]
{Fourth-order} &
$6'/m$, $6'/mmm'$, $\bar{4}3m$, $m'\bar{3}'m$ \\
\hline
\hline
\end{tabular}
\end{table}
\end{center}

\bigskip
\textit{Symmetry analysis}.---
Next, we investigate what kind of symmetry will host a non-vanishing third-order intrinsic Hall current
and when this response becomes the leading contribution.
It has been well known that the number of independent components of a physical quantity such as conductivity tensor
is dictated by the magnetic point group (MPG) symmetry of the system, as required by Neumann's principle
\cite{Neumann,anisotropy}.
Under $\mathcal{T}$-symmetry, the field dependent BC is $\mathcal{T}$-odd and
hence there is no intrinsic Hall signal in $\mathcal{T}$-invariant systems.
However, for materials with $\mathcal{T}$-broken MPGs,
the BC is nonzero and hence we can observe the intrinsic Hall signal in these systems.

In particular, for the rank-4 IAHE conductivity tensor $\chi_{\alpha\beta\gamma\eta}^{\text{int}}$,
the constraint imposed by MPG symmetry operations $R$ and $R\mathcal{T}$ can be expressed as \cite{anisotropy}:
\begin{align}
\chi^{\text{int}}_{\alpha\beta\gamma\eta}
=
\eta_T
R_{\alpha\alpha'}
R_{\beta\beta'}
R_{\gamma\gamma'}
R_{\eta\eta'}
\chi^{\text{int}}_{\alpha'\beta'\gamma'\eta'}
\label{symmetry}
\end{align}
where $\eta_T=1 ~ (\eta_T=-1)$ is for $R(R\mathcal{T})$ and
$R_{\alpha\alpha'}$ is the matrix element of the spatial point group operation $R$.
Starting from Eq.(\ref{symmetry}), in principle we can classify the $90$ 3D MPGs without $\mathcal{T}$
(ruled out $32$ grey MPGs in total $122$ MPGs in 3D) using the 
linear, second-order,  third-order, and other high-order IAHEs.
For example, considering the MPG $6'$ with the generator $C_6\mathcal{T}$, we find
$\chi^{\text{int}}_{xzxx}=-\chi^{\text{int}}_{zxxx}=\chi^{\text{int}}_{zxyy}=-\chi^{\text{int}}_{xzyy}
=-\chi^{\text{int}}_{yzxy}=-\chi^{\text{int}}_{yzyx}=\chi^{\text{int}}_{zyxy}=\chi^{\text{int}}_{zyyx} \neq 0$
and
$\chi^{\text{int}}_{xzxy}=-\chi^{\text{int}}_{zxxy}=\chi^{\text{int}}_{xzyx}=-\chi^{\text{int}}_{zxyx} 
=\chi^{\text{int}}_{yzxx}=-\chi^{\text{int}}_{yzyy}=-\chi^{\text{int}}_{zyxx}=\chi^{\text{int}}_{zyyy}\neq 0$,
whereas both the rank-$2$ and rank-3 conductivity tensors $\chi^{\text{int}}_{\alpha\beta}$ and
$\chi^{\text{int}}_{\alpha\beta\gamma}$ vanish for linear IAHE and second-order IAHE, respectively.

On the other hand, we can also define Jahn's notations
\cite{Jahn} $a\{V^2\}$, $a\{V^2\}V$, $a\{V^2\}[V^2]$ and $a\{V^2\}[V^3]$ for linear,
second-order, third-order, and fourth-order IAHE conductivity tensors, respectively,
and then use Bilbao Crystallographic Server \cite{Bilbao} to
find the 3D MPGs with a non-vanishing Hall signal \cite{supplemental}
and identify the leading contribution, as shown in TABLE I.
We find that the linear, second-order,
third-order, and fourth-order IAHEs are supported by
$31$, $39$, $15$, and $4$ 3D MPGs, respectively, which means that
one can observe Hall signal in almost all of $\mathcal{T}$-broken crystals
(The leading contribution for linear and second-order IAHEs can be found in TABLE I of 
Supplemental Material \cite{supplemental},
in which a full clsssifications rather than leading order is also given.)
It should be noted that the MPGs hosting linear/nonlinear IAHE,
are/aren't compatible with ferromagnetism,
therefore, the high-order Hall current response, as the leading contribution,
can only be observed in AFM materials.
We also note that the linear IAHE is normally assumed to be proportional to magnetization such as in FM metals,
but theoretical predictions and experimental observations have recognized that large Hall effects
can also occur in non-collinear or collinear AFM crystals \cite{Chenhua, Christoph, Satoru, Suzuki},
offered by neither of the global magnetic-dipole symmetry-breaking mechanisms \cite{SciAdv, naturereview}.
The symmetry arguments to search for the third-order IAHE in $\mathcal{T}$-broken systems,
as the leading contribution, is our second main result.

\bigskip
\textit{Two-band model}.---
In this section, we employ a two-band model to illustrate our general theory.
To that purpose, let us first consider the typical transport platform with planar geometry,
in which the applied electric field $\vect{E}$ and the generated current both within the plane
(denoted as $xy$ plane without loss of generality),
and assuming the $\vect{E}$ field forming an angle $\theta$ with the principal axis $x$ of the crystal,
namely
$\vect{E}=E(\cos\theta,\sin\theta)$,
we find that the in-plane intrinsic third-order anomalous Hall current density
can be calculated as \cite{note5}:
\begin{align}
J^{(3)}_{\text{AH}}
&=
E^3
\left[
\chi^{\text{int}}_{yxxx}\cos^2\theta
-
\chi^{\text{int}}_{xyyy}\sin^2\theta
\right]
\nonumber \\
&+
\dfrac{1}{2}E^3
\sin{2\theta}
[
\chi^{\text{int}}_{yxxy}
+
\chi^{\text{int}}_{yxyx}
]
\end{align}
Meanwhile, we note that the in-plane intrinsic third-order parallel current $J^{(3)}_{||}$ is zero \cite{note5}.
For simplicity, below we will take $\theta=0$ and
hence $J_{\text{AH}}^{(3)}=\chi_{yxxx}^{\text{int}}E^3$.

\begin{figure}[ht!]
\centering
\includegraphics[width=0.9\columnwidth]{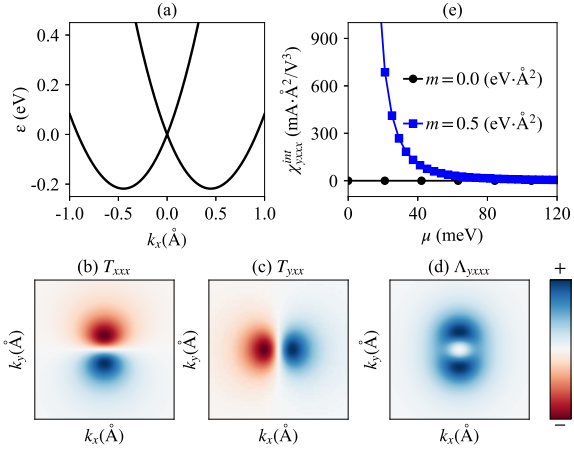}
\caption{
\label{FIG1}
(a) The band structure along $k_x$ direction for the two-band model.
(b-d) The $\vect{k}$-resolved distribution of $T_{xxx}$, $T_{yxx}$, and $\Lambda_{yxxx}$
for the lower band of the two-band model.
(e) The intrinsic third-order anomalous Hall conductivity $\chi^{\text{int}}_{yxxx}$
versus the chemical potential $\mu$.
Model parameters: $t=1.2 (\text{eV}\cdot\mathrm{\AA}^2), v=1.0 (\text{eV}\cdot\mathrm{\AA})$,
and $m=0.5 (\text{eV}\cdot\mathrm{\AA}^2)$,
which approximately describes the valence band edge of monolayer SrMnBi$_2$ around the
$\Gamma$ point\cite{zhang2020higherorder}.}
\end{figure}

Following the full classification in the Supplemental Material\cite{supplemental},
the two-band model respecting MPG $4'm'm$
\begin{align}
H(\vect{k})
=
tk^2+v(k_y\sigma_x-k_x\sigma_y)+m(k_x^2-k_y^2)\sigma_z
\label{model}
\end{align}
is chosen to demonstrate the third-order IAHE, which is a low energy effective Hamiltonian
to describe the band behavior around the $\Gamma$ point of monolayer SrMnBi$_2$ \cite{zhang2020higherorder}.
In Eq.(\ref{model}), $\sigma_i (i=x,y,z)$ represent the Pauli matrices acting on the spin,
and $k^2=k_x^2+k_y^2$. The band dispersion for this model
is $\epsilon_{\pm}=tk^2 \pm h$ with $h^2=v^2k^2+m^2(k_x^2-k_y^2)^2$ even in $k_x$ and $k_y$,
as shown in FIG.\ref{FIG1}(a),
which resembles the Rashba dispersion but with a $\mathcal{T}$-broken second-order warping term
\cite{zhang2020higherorder}.
Interestingly, this effective model also has been utilized to
domonstrate the existence of the third-order extrinsic anomalous Hall effect,
as the leading contribution induced by Berry quadrupole \cite{zhang2020higherorder}.

For this model, the conventional BC for the lower band is found to be:
\begin{align}
\Omega_{z}=-\dfrac{mv^2(k_x^2-k_y^2)}{2h^3}
\end{align}
which is even in $k_x$ and $k_y$. However, due to the mirror symmetry $\mathcal{M}_{x+y}$,
the integral for $\Omega_z$ in the Brillouin zone
vanishes and hence there is no linear IAHE.
Similarly, the first-order field-dependent BC for the lower band is given as:
\begin{align}
\Omega^{(1)}_z/E_x =
-\dfrac{v^2k_y\left[ v^2+m^2(3k_x^2+k_y^2) \right]}{2h^5}
\end{align}
which is odd in $k_y$ and hence makes no contribution to IAHE at the second order.
Note that this result is not inconsistent with our full classification,
in which the system with MPG $4'm'm$
can hold a leading-order second-order IAHE
but involving an out-of-plane index,
in which the nonvanishing components for the second-order IAHE conductivity tensor will be
$\chi_{xzy}^{\text{int}}=\chi_{yzx}^{\text{int}}=-\chi_{zyx}^{\text{int}}=-\chi_{zxy}^{\text{int}}$.
Here we focus on the third-order nonlinear transport behavior of this model.
In terms of our third-order semiclassical theory,
the second-order field-dependent BC for the lower band
can be calculated as: 
\begin{align}
\Omega^{(2)}_z/E_x^2
=
\partial_x T_{yxx}-\partial_y T_{xxx}
=
\dfrac{g_1m^3v^2+g_2mv^4}{8h^{7}}
\label{secondBPT}
\end{align}
with $T_{yxx/xxx}$ the second-order BPTs and $g_{1/2}$ even functions of $k_x$ and $k_y$ \cite{note6}.
Therefore, we conclude that the second-order field-dependent BC will
lead to a nonzero third-order Hall current response.
More intuitively, we present the $\vect{k}$-resolved distribution for
$T_{xxx}$, $T_{yxx}$, and $\Lambda_{yxxx}$ (same as the $\Omega^{(2)}_z/E_x^2$)
for the lower band, as shown in FIG.\ref{FIG1}(b-d),
which determine the final result of $\chi_{yxxx}^{\text{int}}$. 
Particularly, we find that both $T_{xxx}$ and $T_{yxx}$
exhibit a dipole pattern in momentum space, respectively,
and the resultant integrand $\Lambda_{yxxx}$ approximately features a negative ellipse
landscape around $\Gamma$ point. In FIG. \ref{FIG1}(e), we calculate the third-order IAHE
conductivity $\chi^{\text{int}}_{yxxx}$ at zero temperature,
as a function of the chemical potential ($\mu>0$ is assumed),
from which we conclude that the third-order response is nonzero and
becomes significant when the chemical potential is close to zero energy,
manifesting the fact that the third-order intrinsic conductivity is concentrated
around the small-gap region \cite{liu2021intrinsic}.
As a check, we find the third-order IAHE conductivity $\chi^{\text{int}}_{yxxx}$
will disappear when $m=0$, namely when the $\mathcal{T}$-symmetry is recovered,
as can also be easily seen from the Eq.(\ref{secondBPT}).

Interestingly, when the chemical potential $\mu$ is around the band crossing point,
we find that the band dispersion can be approximated as $\epsilon_{\pm}=\pm vk$ and
the second-order BPT can be approximated as $T_{xxx}^{\pm}=\pm 3mk_y/(8v^3k^5)$
and $T_{yxx}^{\pm}=\mp mk_x/(4v^3k^5)$, where $\pm$ denote the conduction band and valence band, respectively,
then the third-order IAHE conductivity at zero temperature can be analytically calculated as 
$\chi_{yxxx}^{\text{int}}=5me^4/(32\pi\hbar|\mu|^3)$\cite{zeroenergy},
which shows a cubic dependence on $|\mu|^{-1}$, consistent with our numerical results.
Note that we have recovered $e$ and $\hbar$ in our final result.
Particularly, when $\mu=0.01$ (\text{eV}), we find that 
$\chi_{yxxx}^{\text{int}} \sim 10^3 (\text{mA}\cdot\mathrm{\AA}^2/\text{V}^3)$,
which corresponds to a Hall voltage $ \sim 0.001 (\mu\text{V})$
by taking the electric field $\sim 10^4 (\text{V}/\text{m})$ \cite{zhang2020higherorder}, 
the resistance $\sim 10^3 (\Omega)$ \cite{zhang2020higherorder} and the lateral size for Hall bar
$\sim 100 (\mathrm{\mu m})$ \cite{liu2021intrinsic}.
Although this Hall voltage is lower by an order of magnitude than the second-order intrinsic anomalous Hall voltage
\cite{liu2021intrinsic},
which can be detected experimentally.

\bigskip
\textit{Summary}.---
In this work, we develop the third-order semiclassical theory
for Bloch electrons under the uniform external electric field
with the semiclassical wavepacket approach.
As one of the important applications,
we predict that the third-order IAHE, driven by the band geometric
quantity---the second-order BC arising from the
second-order field-induced \textit{positional shift}, can occur in $\mathcal{T}$-broken systems.
It should be emphasized that the third-order IAHE, as an important member of
the nonlinear Hall family, has not been explored so far due to the lack of an appropriate theoretical approach.
Furthermore, with symmetry arguments, we find that almost all the MPGs without time-reversal symmetry
can be classified by linear, second-order, third-order, and also fourth-order IAHEs.
Importantly, we find the intrinsic third-order nonlinear anomalous Hall signal, as the
leading contribution, can be accommodated by $15$ 3D MPGs and hence fill the gap in previous studies,
especially in AFM spintronics.
Finally, a two-band toy model is employed to demonstrate the generalized theory.

Last but not least, we note that our third-order semiclassical theory
relies only on the properties of Bloch bands, 
which indicates that our theory can be combined with
first-principles calculations to explore the IAHE of the realistic AFM materials.
For example, following our symmetry analysis,
the AFM materials MnTe \cite{MnTe} and CoNb$_3$S$_6$ \cite{CoNb3S6} should exhibit
a leading-order three-order IAHE signal, which will be explored in future works.

\section*{Acknowledgements}
This work was financially supported by the Natural Science Foundation of China
(Grant No.12034014 and No. 12004442) and
Guangdong Basic and Applied Basic Research Foundation (Grants No. 2021B1515130007).


\begin{thebibliography}{00}
\bibitem{Neumann} Neumann's principle can be stated as:
if a crystal is invariant with respect to certain symmetry,
any of its physical properties must also be invariant with respect to the same symmetry.
As a consequence of this principle, symmetry imposes constraint on the physical quantity,
making the relevant effect disappear.
\bibitem{note6}
We have 
$T_{yxx}=k_xmv^2/(4h^5)$
and 
$T_{xxx}=-3 k_ymv^2/(8h^5)$
from which we obtain 
$g_1=-15 k_x^4 + 40 k_x^2 k_y^2 - 25 k_y^4$
and 
$g_2=-5 k_x^2 - 10 k_y^2$.
Since $g_2$ is negative definite,
its contribution to Eq.(\ref{secondBPT}) is nonzero and hence the third-order IAHE is nonzero for this model.
\bibitem{zeroenergy}
We note that the final result $\chi_{yxxx}^{\text{int}}=5me^4/(32\pi\hbar|\mu|^3)$ 
diverges when the chemical potential $\mu$ lies exactly at the band crossing point,
which may be regularized by disorder effect \cite{wang2021intrinsic}.
\end{thebibliography}
\end{document}